# Low power consumption grating magneto-optical trap based on planar elements

Zhilong Yu[1,†], Yumeng Zhu[1,†], Minghao Yao[1], Feng Qi[1], Liang Chen[2], Chang-ling Zou[2], Junyi Duan[3], and Xiaochi Liu[1,*]

[1]Key Laboratory of Atomic Frequency Standards, Innovation Academy for Precision Measurement Science and Technology, Chinese Academy of Sciences, Wuhan, 430071, China

[2] Chinese Academy of Sciences Key Laboratory of Quantum Information, University of Science and Technology of China, Hefei 230026, China

[3]Center for Advanced Measurement Science, National Institute of Metrology, Beijing, 100029, China

[†]: contributed equally to this work

[*] liuxc@apm.ac.cn

**Abstract**：The grating-based magneto-optical trap (GMOT) is a promising approach for miniaturizing cold-atom systems. However, the power consumption of a GMOT system dominates its feasibility in practical applications. In this study, we demonstrated a GMOT system based on planar elements that can operate with low power consumption. A high-diffraction-efficiency grating chip was used to cool atoms with a single incident beam. A planar coil chip was designed and fabricated with a low power consumption nested architecture. The grating and coil chips were adapted to a passive pump vacuum chamber, and up to $10^6$ $^{87}$Rb atoms were trapped. These elements effectively reduce the power consumption of the GMOT and have great potential for applications in practical cold-atom-based devices.

## 1. Introduction

The magneto-optical trap (MOT) is a general and robust approach for cooling and trapping atoms and is widely applied in quantum standards and sensors [1-7]. However, the application of conventional MOT in mobile and compact systems is still challenging because of their relatively complicated and large architecture. Over the past few years, various efforts have been devoted to simplifying and miniaturizing different elements of the MOT system. The conventional six cooling beams are reduced to a single input beam using a planar grating chip [8-11]. A similar approach can also be realized with a metasurface chip combined with additional optical components [12-14]. Recently, a planar coil chip design was proposed to replace a pair of anti-Helmholtz coils, significantly reducing the volume of the quadrupole magnetic field system [15,16]. The atoms should be trapped in a high-level vacuum system. Thus, different compact and miniaturized vacuum chambers have also been developed [17-20]. These grating-based MOT (GMOT) systems have been verified for their feasibility in several quantum systems, such as atomic clocks and atomic interferometers [21-24].

Previous studies have focused on the miniaturization of the MOT system. Nevertheless, the power consumption of the GMOT systems is noteworthy for practical applications. In this study, we demonstrated a GMOT system based on planar elements with low power consumption. The entire system consists of a high-diffraction-efficiency grating chip, planar nested-coil chip, and a passively pumped compact vacuum chamber. All the core components feature miniaturization and low power consumption. The high diffraction efficiency of the grating chip increases the optical power utilization rate of a single incident cooling beam. The nested design of the proposed planar coil chip significantly reduced power consumption and heating temperature. The non-evaporable getter (NEG) integrated into the vacuum chamber maintains the vacuum level for a long time without an active pump or ion pump. Up to $10^6$ $^{87}$Rb atoms could be cooled and trapped in the GMOT system.

## 2. Microfabrication of the grating chip

A deep-etching grating chip coated with a 100 nm gold film was used to cool the atoms with a single input cooling beam. Fig. 1(a) illustrates the basic principle of atom cooling using a grating atom chip. The incident cooling laser and three diffracted beams form a balanced optical scattering force. Fig. 1(b) and Fig. 1(c) show photographs and scanning electron microscopy (SEM) images of the grating chip. The total chip size was 2 × 2 cm. The grating period was 1400 nm with a grid line interval of 700.1 nm. The efficiency of 1st-order diffracted beam achieved up to 41 %.

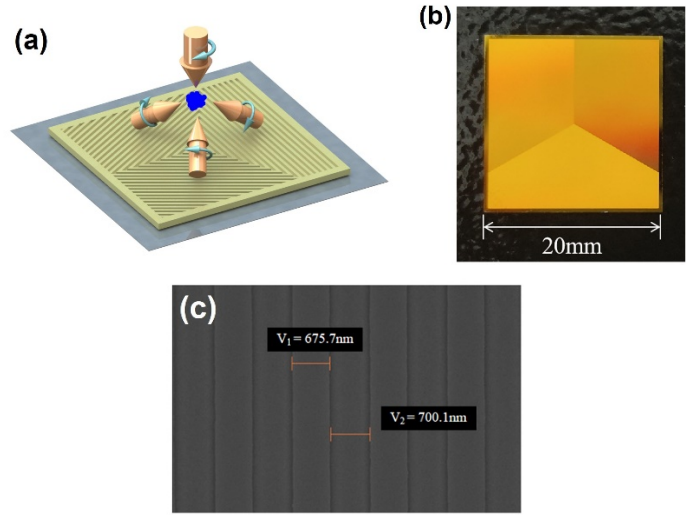

**Fig. 1.** (a) Basic principle of cooling atoms with a grating atom chip; (b) Photo of the grating chip; (c) Partial SEM image of the grating chip.

## 3. Design and simulation of nested anti-Helmholtz coil

In the MOT system, the atoms were slowed down by a cooling laser and trapped in a quadrupole magnetic field with a strength gradient of up to 10 Gs/cm. A planar coil chip was demonstrated to replace a conventional anti-Helmholtz coil pair [16]. The basic idea of this design is to use two groups of coils with opposite current directions to generate a null magnetic field strength at the center with a sufficient gradient, as shown in Fig. 2(a). The magnetic field strength and gradient were calculated as follows [25]:

$$B_z(0,0,z_0) = \frac{\mu_0}{2}\left[\frac{R_1^2 I_1}{\left(R_1^2+z_0^2\right)^{\frac{3}{2}}} - \frac{R_2^2 I_1}{\left(R_2^2+z_0^2\right)^{\frac{3}{2}}}\right], \quad (1)$$

$$\frac{\partial B_z}{\partial z}(0,0,z_0) = -\frac{3\mu_0}{2}\left[\frac{R_1^2 I_1}{\left(R_1^2+z_0^2\right)^{\frac{5}{2}}} - \frac{R_2^2 I_2}{\left(R_2^2+z_0^2\right)^{\frac{5}{2}}}\right]. \quad (2)$$

where $\mu_0$ is the vacuum permeability ($\mu_0 = 4\pi \times 10^{-7}$ N/A$^2$), $I$ is the driving current, $z_0$ is the working point, and $R_1$ and $R_2$ are the various coil radii.

The null magnetic field at $z_0$ position is achieved by the opposite magnetic fields generated by the inner and outer groups of the coils. Because of its larger radius, the outer group of coils requires a higher current to counteract the magnetic field generated by the inner group. Consequently, the outer-group coils have relatively high-power consumption.

Herein, we proposed a nested planar coil chip design with a careful calculation of the magnetic field strength generated by each wire circle. The current direction for each circle was set individually rather than simply divided into two groups. Eqs. 1 and 2 can be expanded as

$$B_z(0,0,z_0) = \frac{\mu_0}{2}\left(\pm\frac{R_0^2 I_0}{\left(R_0^2+z_0^2\right)^{\frac{3}{2}}} \pm \frac{R_1^2 I_0}{\left(R_1^2+z_0^2\right)^{\frac{3}{2}}} \cdots\cdots \pm \frac{R_i^2 I_0}{\left(R_i^2+z_0^2\right)^{\frac{3}{2}}}\right), \quad (3)$$

$$\frac{\partial B_z}{\partial z}(0,0,z_0) = \frac{3\mu_0}{2}\left(\pm\frac{R_0^2 I_0}{\left(R_0^2+z_0^2\right)^{\frac{5}{2}}} \pm \frac{R_1^2 I_0}{\left(R_1^2+z_0^2\right)^{\frac{5}{2}}} \cdots\cdots \pm \frac{R_i^2 I_0}{\left(R_i^2+z_0^2\right)^{\frac{5}{2}}}\right). \quad (4)$$

where $I_0$ is the driving current and $R_n$ is the coil radius. Fig. 2(b) shows an example of a nested coil chip in which the coils are divided into three groups.

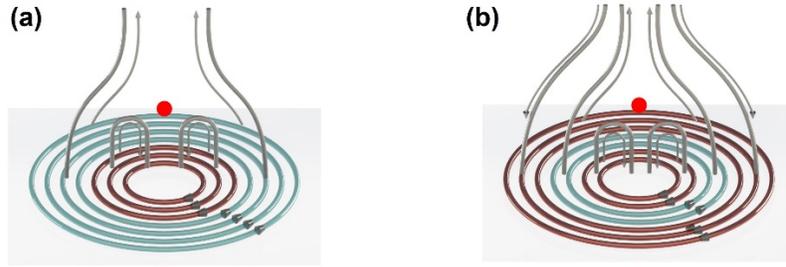

**Fig. 2.** Planar anti-Helmholtz coil chip structure. (a) Two groups coils; (b) Three groups nested coils.

Fig. 3 plots the theoretical calculated results of the power consumption P of the nested coils design versus the height of the null magnetic field h, the circle number N, and the layer number, respectively. The data plotted in Fig. 3 all satisfy two constraints: the magnetic field strength is zero and the gradient is 11. 86 Gs/cm. The points in Fig. 3 are calculated using Eqs. 3 and 4. Several approximate design results could be obtained for each point. We selected the design corresponds the lowest P. For example, four different designs are obtained when N=25 shown in the inset image of Fig. 3(b). The red circle presents the clockwise current direction, and the current of the green circles are inverse. The power consumptions calculated of these four designs are 697.3, 808.4, 742.5, and 877.2 mW, respectively. Among these designs, although the power consumptions are approximate, the lowest one (the 1st design in this case) is selected for the data plotted in Fig. 3.

In Fig. 3(a), the number of circles *N* was fixed at 25. The power consumption *P* exponentially increases with the height of the null magnetic field, *h*. In the GMOT system, the center of the diffracted beams overlap of the grating chip must overlap with the center of the quadrupole magnetic field. Thus, the operating height of the coil chip design was set to 6.5 mm, which is equal to the height of the center of the beam overlap. Based on this fixed height, Fig. 3(b) presents the power consumption as a function of the number *of circles N*. The power consumption is decreased with more circle on the coils chip. However, the size of the planar chip is limited to the total dimensions of the GMOT system. The circle number of the coil chip design was selected as 25 to adapt to the grating chip and vacuum chamber. The relationship between the number of overlapped coil layers and the power consumption is presented in Fig. 3(c) using the first coil chip design shown in Fig. 3(b). The power consumption of the coil decreased with an increase in the number of overlapping layers.

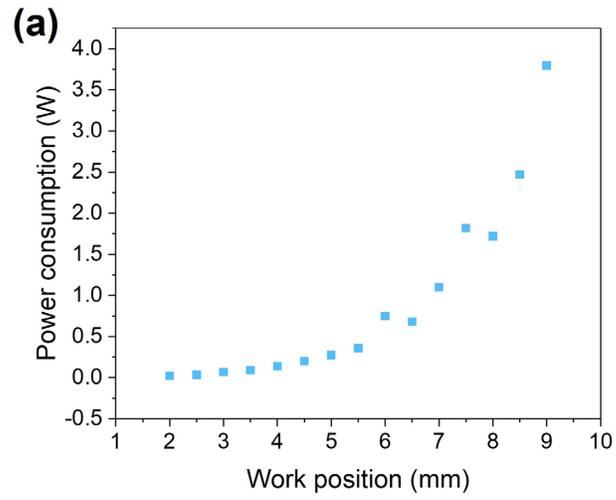

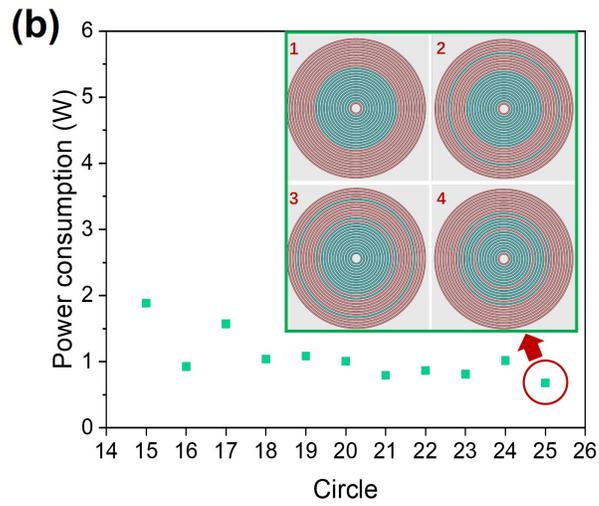

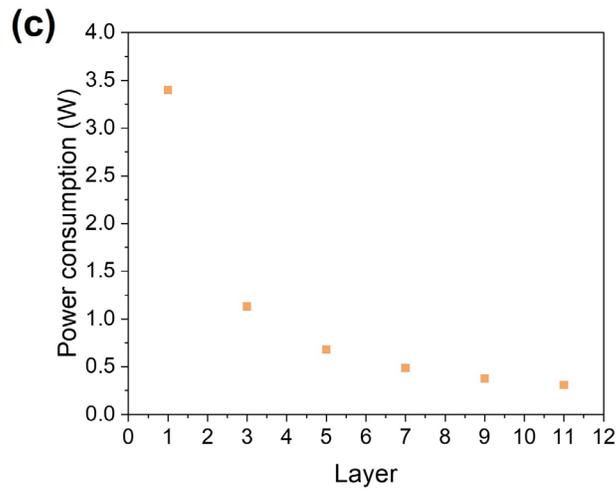

**Fig. 3.** (a) Null magnetic field height versus power consumption; (b) Circle number versus power consumption; (c) Coils layer number versus power consumption.

## 4. Test results nested anti-Helmholtz coil

Based on the simulation results, a coil chip with the design shown in Fig. 3(c) was fabricated using a printed circuit board. A photograph of the chip is shown in the inset of Fig. 4(a). The physical model had a thickness of approximately 1 mm and an area of 39 × 39 mm. It comprises five parallel copper wires, each layer having a thickness of 70 $\mu$m, a width of 545 $\mu$m, and a wire spacing of 150 $\mu$m. In each layer, the innermost coil has $N_1 = 1$ and the radius is 1.4 mm, the middle coil has $N_2 = 13$, with a radius range of 2.095–10.435 mm, and the outermost coil has $N_3 = 11$, with a radius range of 11.13–18.08 mm. Each layer of the coil was interconnected through metal vias and connected to bottom-printed pads on the chip.

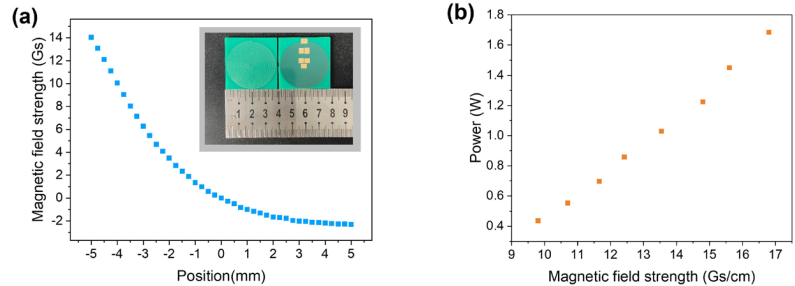

**Fig. 4.** (a) Measured magnetic field strength versus operating height; (b) Magnetic field gradient versus power consumption.

The measured performance of the nested coil chip is shown in Fig. 4(a) and Fig. 4(b). A Gaussian meter is used to measure the strength of the magnetic field along the coil axis. The current was set to 2.2 A for the measurements in Fig. 4(a), resulting in a magnetic field gradient of 12.4 Gs/cm. The magnetic-field strength was nullified at an operating height of 6.5 mm. The steady-state magnetic field gradient of the coil chip and the corresponding power consumption were within the range required by the magneto-optical trap system, as depicted in Fig. 4(b). At the desired gradient of 12.4 Gs/cm, the applied current was 2.2 A, resulting in a power consumption of 0.825 W.

## 5. Compact vacuum system

Another key component was the vacuum chamber. The basic architecture and a photograph of the vacuum chamber with the grating and coil chips are shown in Fig. 5(a) and Fig. 5(b). The size of the entire chamber was 3 × 3 × 7 cm with five windows of diameter 2 cm. We utilized NEGs to maintain an ultrahigh vacuum and avoid the extra volume and power consumption caused by active pumps [26]. The getters and dispensers were housed in the side chamber. Given that NEGs do not extract rare gases, helium permeation emerges as a critical factor influencing vacuum longevity [27]. We chose titanium as the chamber material, which has low outgassing and permeability [28]. The windows are made of sapphire, which has extensive optical transparency and minimal helium diffusion [29]. Considering the activation and operating temperature range of the getters (SAES St172/HI/7.5-7), we employed the connection between titanium and sapphire proves robust, theoretically endowing the chamber with excellent sealing performance. The ion pump and vacuum gauge are connected to the chamber through a copper tube for initial vacuum pumping and pressure monitoring.

The pressure must be measured while independently operating to characterize the vacuum performance of the chamber. We employed the loading time of GMOT as a means to assess the efficacy of our vacuum system [30-32]. The loading dynamics of the MOT can serve as a reliable gauge of vacuum pressure, with loading time indicating a pressure at or below $(2 \times 10^{-8}\,\text{Torr} \cdot \text{s})/\tau$ [31]. This not only showcases the chamber's suitability for cold-atom experiments but also provides insights into the longitudinal evolution of the background pressure. Despite the influence of trap parameters [31], this provided a reliable upper limit for the background pressure of the vacuum system. Fig. 5(c) shows the long-term vacuum level monitoring up to 1500 h. It is evident that the pressure within the chamber remains consistently below $5\times10^{-8}$ Torr, displaying no discernible trend of deterioration.

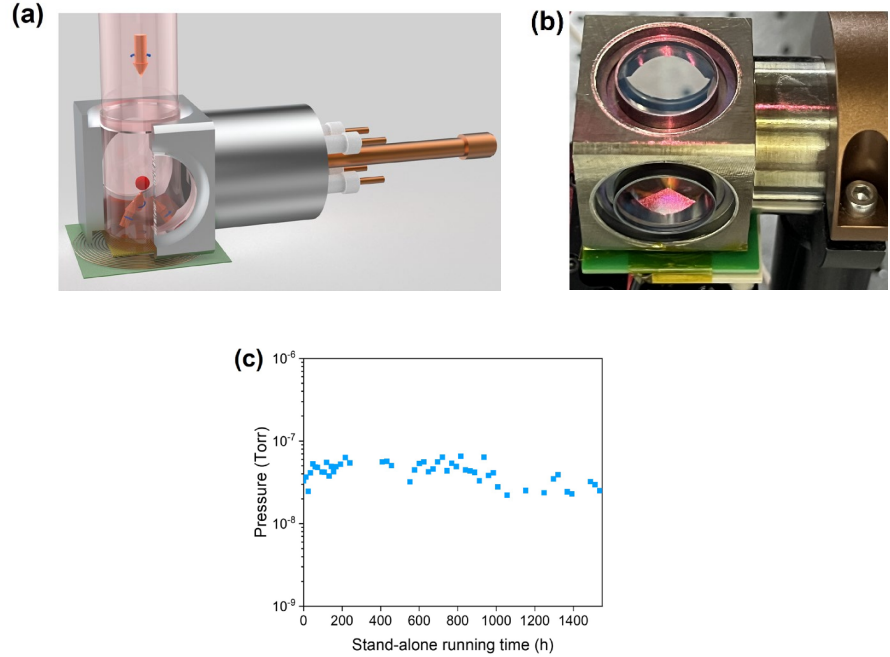

**Fig. 5.** (a) Basic architecture of the fabricated vacuum chamber with the grating chip and the coils chip; (b) Photo of the fabricated vacuum chamber; (c) Measured long term vacuum level.

To verify the performance of the aforementioned core components, we constructed a GMOT system for the cold-atom experiments. A grating chip was mounted underneath and outside the chamber, as Fig. 5(b) shown. The coil chip was installed and aligned below the grating chip, such that the working areas of the two chips coincided. The side windows facilitated atomic image observation and fluorescence detection. The light sources used for the experiment were two distributed Bragg reflector lasers. The 780 nm cooling laser frequency was tuned to the $5S_{1/2}(F=2)$ to $5P_{3/2}(F'=3)$ cycling transition with a red detuning of approximately 8.8 MHz, and the 795 nm repump laser was tuned to the $5S_{1/2}(F=1)$ to $5P_{3/2}(F'=2)$ transition. The beams of the two lasers were combined and coupled into a polarization-maintaining fiber, and delivered to the MOT setup. The intensity of the incident beam is 7 mW/cm$^2$.

## 6. Experimental results

Fig. 6(a) shows an image of the cold atoms captured by the CCD camera. The fluorescence detection results indicate that up to $1.6 \times 10^6$ cold atoms can be trapped by 20 mW laser power incident on the grating chip with a surface of $20 \times 20$ mm. The number of atoms trapped in the GMOT depends on the cooling laser detuning and the optical power, as shown in Fig. 6(b) and Fig. 6(c). Moreover, the number trapped reached the maximum value when the cooling laser detuning was approximately 9 MHz. Within the parameter range, the number of atoms exhibited an approximately linear relationship with the laser power. In addition, we measured the number of atoms at different currents (corresponding to power consumption) to verify the performance of the coil chip, as shown in Fig. 6(d). The results indicate that the nested coil chip can laser cool and trap $10^6$ cold atoms at a lower power consumption. This number increases as the coil power consumption increases.

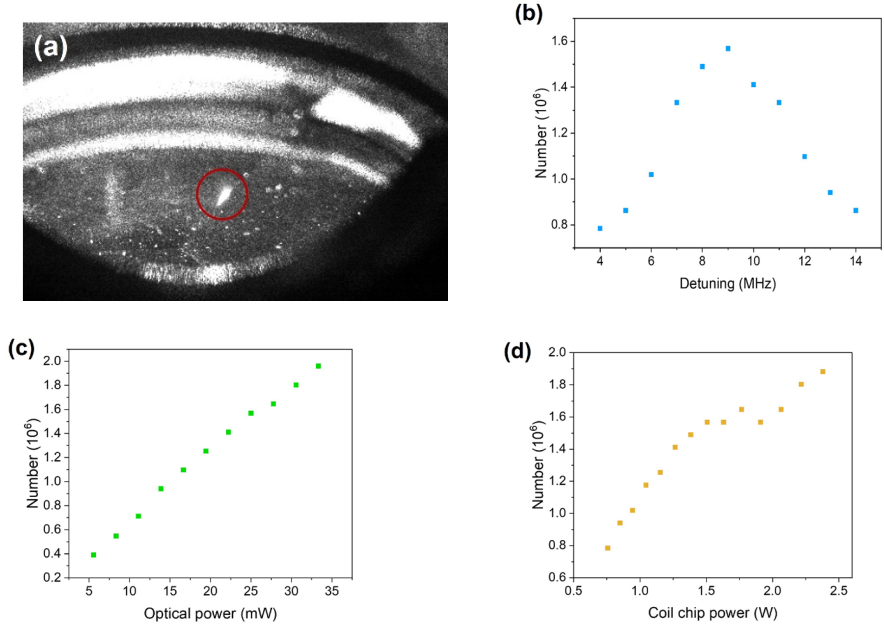

**Fig. 6.** (a) Cold atoms fluorescence marked with the red circle; (b) Cold atom number versus the cooling laser detuning; (c) Cold atom number versus laser power; (d) Cold atom number versus laser power coil chip power consumption.

## 7. Conclusions and discussion

We demonstrated a low-power-consumption GMOT system based on planar elements. The nested configuration significantly reduces the power requirements and heat dissipation of the coil chip. The chamber can sustain an ultrahigh vacuum for laser cooling with passive pumping for a long time. Combined with a diffraction efficiency up to 40% planar grating chips, up to $10^6$ cold atoms were trapped. The results demonstrate impressive progress toward reducing the power consumption and volume of a compact cold-atom system. It holds great potential for application in portable cold-atom sensors and quantum devices [33-37]. Several interesting points must be discussed.

Owing to the periodic structure of the grating chip, the scattering forces of the incident and diffracted light can be better balanced only when the light incident on its surface is as uniform as possible. In a GMOT system, Gaussian light emitted from an optical collimator is expanded by a lens to obtain a flat distribution beam [8,19,38]. This configuration requires high optical power to guarantee sufficient effective optical intensity for laser cooling. A refractive optical system has also been proposed that converts a Gaussian beam to a flat-top beam [40]. However, this method increases the complexity of the system, which hinders miniaturization. We first attempted to implement this function using commercial diffractive optics to reduce the laser power used in our system. The cooling laser power can be reduced to 25 % of the Gaussian case with a similar experimental result. A metasurface is an effective approach for converting a Gaussian beam into a flat-top beam by maintaining the compact architecture of the whole system [40-42].

The coil chip we designed works with a surface temperature that ultimately stabilizes at around 45 ℃, which is 30 ℃ lower than the two groups coils structure [Fig. 2(a)]. However, for practical applications, the coil temperature must be further improved. The selection of wires and substrate materials with low resistance and high conductivity can reduce the internal resistance and energy loss of the coil. Processing more layers to improve the heat dissipation efficiency is also feasible. In addition, the height of the operating point was 6.5 mm, which increased the power consumption and temperature, according to the calculated results in Fig. 3a. If a new grating chip design can lower the height of the beam overlap center, the power consumption of the coil chip can be decreased. The magnetic coil chips and grating chips in the GMOT system can be fabricated on the same silicon wafer to further simplify the structure and reduce the number of devices required. Note that the conventional structure coils may achieve lower power consumption than the nested planar design, but for the laser cooling technique on chip system, the planar elements are more preferred than the cubic structure. We have also simulated the nested coils chip with the same size of the previous work [16], the power consumption is still lower. With larger surface of the coils chip, the advantage of the nested design is more obvious.

The vacuum chamber used in our system was fabricated using Ti and sapphire. The birefringent crystalline structure of sapphire can be overcome by being cut into special shapes such that a single optical axis will be unaffected [18]. We can obtain the maximum number of cold atoms by adjusting the polarization of the input beam in the experiment. But it is difficult to further reduce the volume of the chamber due to limitations in processing methods. Other solutions, such as a micro-electromechanical-systems (MEMS) cells, may be a good choice for improving the vacuum system [17]. The difficulty of fluorescence detection also needs to be solved through structural adjustment [43]. This approach might be an ideal candidate for laser cooling in low-SWaP applications in the future if long-term vacuum retention capability is achieved.

**Funding.** National Natural Science Foundation (12273087); National Key Research and Development Program of China (2021YFF0603701).
**Disclosures.** The authors declare no conflicts of interest.
**Data availability.** Data underlying the results presented in this paper are not publicly available at this time but may be obtained from the authors upon reasonable request.

**References**

1. J. Grotti, S. Koller, S. Vogt, S. Hafner, U. Sterr, C. Lisdat, H. Denker, C. Voigt, L. Timmen, Roll, F.N. Baynes, H.S. Margolis, M. Zampaolo, P. Thoumany, M. Pizzocaro, B. Rauf, F. Bregolin, A. Tampellini, P. Barbieri, M. Zucco, G.A. Costanzo, C. Clivati, F. Levi, and D. Calonico, "Geodesy and metrology with a transportable optical clock," Nat. Phys. 14(5), 437-441 (2018).
2. T. Udem, T. Holzwarth, and T.W. Hansch, "Optical frequency metrology," Nature. 416(6877), 233-237 (2002).
3. M. Schioppo, R. C. Brown, W. F. McGrew, N. Hinkley, R. J. Fasano, K. Beloy, T. H. Yoon, G. Milani, D. Nicolodi, J. A. Sherman, N. B. Phillips, C. W. Oates, and A. D. Ludlow, "Ultrastable optical clock with two cold-atom ensembles," Nat. Photon. 11(1), 48-52 (2017).
4. S. Bize, P. Laurent, M. Abgrall, H. Marion, I. Maksimovic, L. Cacciapuoti, J. Grünert, C. Vian, F. P. dos Santos, P. Rosenbusch, P. Lemonde, G. Santarelli, P. Wolf, A. Clairon, A. Luiten, M. Tobar, and C. Salomon, "Cold atom clocks and applications," J. Phys. B: At. Mol. Opt. Phys. 38(9), S449 (2005).
5. L. Liu, D. Lu, W. Chen, T. Li, Q. Qu, B. Wang, L. Li, W. Ren, Z. Dong, J. Zhao, W. Xia, X. Zhao, J. Ji, M. Ye, Y. Sun, Y. Yao, D. Song, Z. Liang, S. Hu, D. Yu, X. Hou, W. Shi, H. Zang, J. Xiang, X. Peng, and Y. Wang, "In-orbit operation of an atomic clock based on laser-cooled 87Rb atoms," Nat. Commun. 9(1), 2760 (2018).
6. R. Geiger, A. Landragin, S. Merlet, and F. Pereira Dos Santos, "High-accuracy inertial measurements with cold-atom sensors,"

AVS Quantum Sci. 2(2), 024702 (2020).

7. K. D. Nelson, C. D. Fertig, P. Hamilton, J. M. Brown, B. Estey, H. Müller, and R. L. Compton, "Guided matter wave inertial sensing in a miniature physics package," Appl. Phys. Lett. 116(23), 234002 (2020).

8. J. Duan, X. Liu, Y. Zhou, X. Xu, L. Chen, C. Zou, Z. Zhu, Z. Yu, N. Ru, and J. Qu, "High diffraction efficiency grating atom chip for magneto-optical trap," Opt. Commun. 513, 128087 (2022).

9. P. McGilligan, P. F. Griffin, E. Riis, and A. S. Arnold, "Diffraction-grating characterization for cold-atom experiments", J. Opt. Soc. Am. B. 33(6), 1271–1277 (2016).

10. J. Kitching, "Chip-scale atomic devices," Appl. Phys. Rev. 5(3), 031302 (2018).

11. C. C. Nshii, M. Vangeleyn, J. P. Cotter, P. F. Griffin, E. A. Hinds, C. N. Ironside, P. See, A. G. Sinclair, E. Riis and A. S. Arnold, "A surface-patterned chip as a strong source of ultracold atoms for quantum technologies," Nat. Nano. 8(5), 321–324 (2013).

12. J. Engelberg and U. Levy, "The advantages of metalenses over diffractive lenses," Nat. Commun. 11(1), 1991 (2020).

13. M. Jin, X. Zhang, X. Liu, C. Liang, J. Liu, Z. Hu, K. Li, G. Wang, Jun Yang, L. Zhu, and G. Li, "A Centimeter-Scale Dielectric Metasurface for the Generation of Cold Atoms," Nano. Lett. 23(9), 4008–4013 (2013).

14. W. R. McGehee, W. Zhu, D. S. Barker, D. Westly, A. Yulaev, N. Klimov, A. Agrawal, S. Eckel, V. Aksyuk and J. J. McClelland, "Magneto-optical trapping using planar optics," New J. Phys. 23(1), 013021 (2021).

15. J. D. Weinstein, K. G. Libbrecht, "Microscopic magnetic traps for neutral atoms," Phys. Rev. A. 52(5), 4004 (1995).

16. L. Chen, C. Huang, X. Xu, Y. Zhang, D. Ma, Z. Lu, Z. Wang, G. Chen, J. Zhang, H. Tang, C. Dong, W. Liu, G. Xiang, G. Guo, and C. Zou, "Planar-Integrated Magneto-Optical Trap," Phys. Rev. Appl. 17(3), 034013 (2022).

17. J. P. McGilligan, K. R. Moore, A. Dellis, G. D. Martinez, E. de Clercq, P. F. Griffin, A. S. Arnold, E. Riis, R. Boudot, and J. Kitching, "Laser cooling in a chip-scale platform," Appl. Phys. Lett. 117(5), 054001 (2020).

18. B. J. Little, G. W. Hoth, J. Christensen, C. Walker, D. J. D. Smet, G. W. Biedermann, J. Lee, and P. D. D. Schwindt, "A passively pumped vacuum package sustaining cold atoms for more than 200 days," AVS Quantum Sci. 3(3), 035001 (2021).

19. O. S. Burrow, P. F. Osborn, E. Boughton, F. Mirando, D. P. Burt, P. F. Griffin, A. S. Arnold, and E. Riis, "Stand-alone vacuum cell for compact ultracold quantum technologies," Appl. Phys. Lett. 119(12), 124002 (2021).

20. N. Cooper, L. A. Coles, S. Everton, I. Maskery, R. P. Campion, S. Madkhaly, C. Morley, J. O'Shea, W. Evans, R. Saint, P. Krüger, F. Oruševiíc, C. Tuck, R.D. Wildman, T. M. Fromhold, and L. Hackermüller, "Additively manufactured ultra-high vacuum chamber for portable quantum technologies," Addit. Manuf. 40, 101898 (2021).

21. R. Elvin, G. W. Hoth, M. Wright, B. Lewis, J. P. McGilligan, A. S. Arnold, P. F. Griffin, and E. Riis, "Cold-atom clock based on a diffractive optic," Opt. Express. 27(26), 38359-38366 (2019).

22. F.-X. Esnault, E. Blanshan, E. N. Ivanov, R. E. Scholten, J. Kitching, and E. A. Donley, "Cold-atom double-λ coherent population trapping clock," Phys. Rev. A. 88(4), 042120 (2013).

23. A. Bregazzi, E. Batori, B. Lewis, C. Affolderbach, G. Mileti, E. Riis, P. Griffin, "A cold-atom Ramsey clock with a low volume physics package," arXiv:2305.02944 [physics. atom-ph]

24. J. Lee, R. Ding, J. Christensen, R. R. Rosenthal, A. Ison, D. P. Gillund, D. Bossert, K. H. Fuerschbach, W. Kindel, P. S. Finnegan, J. R. Wendt, M. Gehl, H. McGuinness, C. A. Walker, A. Lentine, S. A. Kemme, G. Biedermann, and P. D. D. Schwindt, "A compact cold-atom interferometer with a high data-rate grating magneto-optical trap and a photonic-integrated-circuit-compatible laser system," Nat. Commun. 13(1), 5131 (2021).

25. T. Bergeman, G. Erez, and H. J. Metcalf, "Magnetostatic trapping fields for neutral atoms," Phys. Rev. A. 35(4), 1535 (1987).

26. D. R. Scherer, D. B. Fenner, and J. M. Hensley, "Characterization of alkali metal dispensers and non-evaporable getter pumps in ultrahigh vacuum systems for cold atomic sensors," J. Vac. Sci. Technol. A. 30(6), 061602 (2012).

27. R. Boudot, J. P. McGilligan, K. R. Moore, V. Maurice, G. D. Martinez, A. Hansen, E. de Clercq, and J. Kitching, "Enhanced

observation time of magneto-optical traps using micro-machined non-evaporable getter pumps," Sci. Rep. 10(1), 16590 (2020).

28. M. Takeda, H. Kurisu, S. Yamamoto, H. Nakagawa, and K. Ishizawa, "Hydrogen outgassing mechanism in titanium materials," Appl. Surf. Sci. 258(4), 1405 (2011).

29. M. Stephens, R. Rhodes, and C. Wieman, "Study of wall coatings for vaporcell laser traps," J. Appl. Phys. 76(6), 3479 (1994).

30. R. W. G. Moore, L. A. Lee, E. A. Findlay, L. Torralbo-Campo, G. D. Bruce, and D. Cassettari, "Measurement of vacuum pressure with a magneto-optical trap: A pressure-rise method," Rev. Sci. Instrum. 86(9), 093108 (2015).

31. T. Arpornthip, C. A. Sackett, and K. J. Hughes, "Vacuum-pressure measurement using a magneto-optical trap," Phys. Rev. A. 85(5), 033420 (2012).

32. J. Scherschligt, J. A. Fedchak, Z. Ahmed, D. S. Barker, K. Douglass, S. Eckel, E. Hanson, J. Hendricks, N. Klimov, T. Purdy, J. Ricker, R. Singh, and J. Stone, "Review article: Quantum-based vacuum metrology at the National Institute of Standards and Technology," J. Vac. Sci. Technol. A. 36(4), 040801 (2018).

33. A. Kuzmich, W. P. Bowen, A. D. Boozer, A. Boca, C. W. Chou, L.-M. Duan, and H. J. Kimble, "Generation of nonclassical photon pairs for scalable quantum communication with atomic ensembles," Nature 423(6941), 731-437 (2003).

34. Z. Hu, B. Sun, X. Duan, M. Zhou, L. Chen, S. Zhan, Q. Zhang, and J. Luo, "Demonstrationn of an ultrahigh-sensitivity atom-interferometry absolute gravimeter," Phys. Rev. A. 88(4), 043610 (2013).

35. M. J. Snadden, J. M. McGuirk, P. Bouyer, K. G. Haritos, and M. A. Kasevich, "Measurement of the Earth's Gravity Gradient with an Atom Interferometer-Based Gravity Gradiometer," Phys. Rev. Lett. 81(5), 971 (1998).

36. D. Savoie, M. Altorio, B. Fang, L. A. Sidorenkov, R. Geiger, and A. Landragin, "Interleaved atom interferometry for high-sensitivity inertial measurements," Sci. Adv. 4(12), 7948 (2018).

37. A. Periwal, E. S. Cooper, P. Kunkel, J. F. Wienand, E. J. Davis, and M. Schleier-Smith, "Programmable interactions and emergent geometry in an array of atom clouds," Nature 600(7890), 630-635 (2021).

38. S. Eckel, D. S. Barker, J. A. Fedchak, N.N. Klimov, E. Norrgard, J. Scherschligt, C. Makrides and E. Tiesinga, "Challenges to miniaturizing cold atom technology for deployable vacuum metrology," Metrologia. 55(5), S182 (2018).

39. J. A. Hoffnagle and C. M. Jefferson, "Design and performance of a refractive optical system that converts a Gaussian to a flattop beam," Appl. Opt. 39(30),5488-5499 (2000).

40. A. Yulaev, W. Zhu, C. Zhang, D. A. Westly, H. J. Lezec, A. Agrawal, and V. Aksyuk, "Metasurface-Integrated Photonic Platform for Versatile Free-Space Beam Projection with Polarization Control," ACS Photonics 6(11), 2902−2909 (2019).

41. B. Walther, C. Helgert, C. Rockstuhl, F. Setzpfandt, F. Eilenberger, E. Kley, F. Lederer, A. Tünnermann, and T. Pertsch, "Photonics: Spatial and Spectral Light Shaping with Metamaterials," Adv. Mate. 24(47), 6251-6253 (2012).

42. L. Cong, N. Xu, J. Gu, R. Singh, J. Han, and W. Zhang, "Highly flexible broadband terahertz metamaterial quarter-wave plate," L. A. Photo. Rev. 8(4), 626-632 (2014).

43. A. Bregazzi, P. F. Griffin, A. S. Arnold, D. P. Burt, G. Martinez, R. Boudot, J. Kitching, E. Riis, and J. P. McGilligan, "A simple imaging solution for chip-scale laser cooling," Appl. Phys. Lett. 119(18), 184002 (2021).